\documentclass[preprint,superscriptaddress,preprintnumbers, amsmath,amssymb,]{revtex4-1}
\usepackage{xcolor}
\usepackage{hyperref}
\hypersetup{
    colorlinks,
    linkcolor={red!50!black},
    citecolor={blue!50!black},
    urlcolor={blue!80!black}
  }
\usepackage{ulem}  
\usepackage[utf8]{inputenc}

\begin{document}
\title{Revisiting proton-proton fusion in chiral effective field theory}
\author{Bijaya Acharya\footnote{Corresponding author.}}\email{bid@ornl.gov}
\affiliation{Physics Division, Oak Ridge National Laboratory, Oak Ridge, TN 37831, USA}

\author{Laura Elisa Marcucci}\email{laura.elisa.marcucci@unipi.it}
\affiliation{\mbox{Department of Physics ``E. Fermi'', University of Pisa, Pisa, I-56127, Italy}}
\affiliation{\mbox{Istituto Nazionale di Fisica Nucleare, Sezione di Pisa, Pisa, I-56127, Italy}}

\author{Lucas Platter}\email{lplatter@utk.edu}
\affiliation{Department of Physics and Astronomy, University of Tennessee, Knoxville, TN 37996, USA}
\affiliation{Physics Division, Oak Ridge National Laboratory, Oak Ridge, TN 37831, USA}
\date{\today}

\begin{abstract}
  We calculate the $S$-factor for proton-proton fusion using chiral
  effective field theory interactions and currents. By performing
  order-by-order calculations with a variety of chiral interactions
  that are regularized and calibrated in different ways, we assess the
  uncertainty in the $S$-factor from the truncation of the effective
  field theory expansion and from the sensitivity of the $S$-factor to
  the short-distance axial current determined from three- and
  four-nucleon observables. We find that
  $S(0)=(4.100\pm0.024\mathrm{(syst)}\pm0.013\mathrm{(stat)}\pm0.008(g_A))\times10^{-23}~\mathrm{MeV\,fm}^2\,,$
  where the three uncertainties arise, respectively, from the
  truncation of the effective field theory expansion, use of the
  two-nucleon axial current fit to few-nucleon observables and
  variation of the axial coupling constant within the recommended
  range. The increased value of $S(0)$ compared to previous calculations is mainly driven by 
  an increase in the recommended value for the axial coupling constant and is in agreement with 
  a recent analysis based on pionless effective field theory.
  
\end{abstract}

\maketitle
\section{Introduction}
\label{sec:intro}

Nuclear reaction rates are among the main sources of systematic
uncertainty in stellar evolution models~\cite{Vinyoles:2016djt}. The
proton-proton ($pp$) fusion reaction is the rate-determining step of
the $pp$ chains that power the Sun and lighter stars. Available
experimental techniques are not able to directly measure the rate of
this process with sufficient precision at energies relevant for
stellar burning, and values predicted by nuclear theory provide
critical inputs to astrophysical
simulations~\cite{Adelberger:2010qa}. More reliable
calculations of this reaction rate can shed further light on the 
inconsistencies in observed solar data such as those that
exist between spectroscopic determinations of solar abundances and
helioseismology~\cite{Bahcall:2004yr,Serenelli:2009yc}. In concert
with improvements in other physics inputs such as radiative opacities,
rigorous constraints on this rate can also help us use solar data as a
powerful probe of new physics~\cite{Suliga_2021}.

The calculations of this process were traditionally performed using
potential models~\cite{Bethe:1938yy, Bahcall:1968wz,
  Kamionkowski:1993fr,Schiavilla:1998je}. Over the last few decades,
methods based on effective field theory (EFT)
techniques~\cite{Epelbaum:2008ga} that allow us to obtain theoretical
predictions with reliable uncertainty estimates, have been
employed. EFTs provide a simplified yet rigorous description of the
process under study using only those degrees of freedom that are
relevant at low energies. The calculations are organized as systematic
expansions in the ratio of the typical momentum scale $Q$ of the
process to a large momentum scale $\Lambda_b$, beyond with the EFT
expansion breaks down. The undetermined parameters---the so-called
low-energy constants (LECs)---that appear up to a given order in this
$Q/\Lambda_b$ expansion are first fixed, {e.g.}, by fitting to
experimental data, and then predictive calculations are performed for
other observables. The first applications of EFT to the $pp$ fusion
process were based on hybrid approaches~\cite{Park:1998wq,Park:2002yp}
that employed wave functions obtained from phenomenological nuclear
potentials along with the nuclear electroweak current operators
derived in chiral EFT, which employs pions and nucleons as dynamical
degrees of freedom. Complete chiral EFT calculations, using potentials
and currents both derived consistently within the chiral EFT
framework, have been carried out, first in
Ref.~\cite{Marcucci:2013tda} and then in
Ref.~\cite{Acharya:2016kfl}. However, as discussed further below,
these calculations require important updates and corrections.

Over the past few decades, a number of studies have also been performed
in pionless EFT, which uses nucleons as the only dynamical degrees of
freedom~\cite{Kong:2000px,Butler:2001jj,Ando:2008va,Chen:2012hm,De-Leon:2022omx,Acharya:2019zil}. At
energies relevant for astrophysics, the process lies well within the
domain of convergence of pionless EFT. In this approach, the $pp$ fusion rate 
can be calculated with a small number of parameters. The uncertainties 
 have traditionally been dominated by the limitation
of experiments and Lattice QCD to sufficiently constrain the LEC
$L_{1,A}$ that parametrizes the short-distance two-body axial
current~\cite{Acharya:2019fij,Davoudi:2021noh}. By performing next-to-leading-order 
calculations of relevant three-body observables, a recent pionless EFT work~\cite{De-Leon:2022omx} 
has quoted the value $S(0)=(4.14\pm0.01\pm0.005\pm0.06)\times10^{-23})~\mathrm{MeV~fm}^2$, where the three uncertainties arise, respectively, from the experimental errors on the nucleon axial coupling and the tritium $\beta$ decay rate, and the theory uncertainty of pionless EFT. 

The goal of this work is to present a state-of-the-art calculation of
the $pp$ fusion cross section in chiral EFT. In addition to correcting~\cite{PhysRevLett.121.049901}
the erroneous treatment of the relationship between the three-nucleon force parameter $c_D$ and the 
two-nucleon axial current parameter $\hat d_R$, we also improve and expand
upon the work of Ref.~\cite{Acharya:2016kfl} by using recently
developed chiral EFT
interactions~\cite{Reinert:2017usi,Wesolowski:2021cni}, 
and by accounting for uncertainty sources not previously
considered in Ref.~\cite{Acharya:2016kfl}.  We also review and update
the work of Ref.~\cite{Marcucci:2013tda} by taking particular care of
the convergence~\cite{Acharya:2016rek} in the expansion basis used to calculate the nuclear
wave functions, as well as the range of the
integration in the axial current matrix element, by also correcting
the $c_D$-$\hat d_R$ relation, and by using the most recent values for
the fundamental constants, which leads to a reassessment of the
Gamow-Teller matrix element of tritium $\beta$-decay. The same nuclear
interaction of Ref.~\cite{Marcucci:2013tda} is implemented, limiting
the present study to the case of cutoff $\Lambda=500$ MeV. This
calculation also allows us to perform a benchmark study between the
approaches used in Ref.~\cite{Acharya:2016kfl} and the one of
Ref.~\cite{Marcucci:2013tda}. The two approaches will be labelled LS
and VM, respectively, since the former used the Lippmann-Schwinger
equation to solve the two-body problem, while the latter used a
variational method~\cite{PhysRevLett.108.052502} as discussed below.

The paper is organized as follows: we outline the theoretical
framework in Section~\ref{sec:sfac}, present our results in
Section~\ref{sec:results}, and conclude with a summary and outlook in
Section~\ref{sec:summary}.

\section{The $S$-factor}
\label{sec:sfac}

The astrophysical $S$-factor $S(E)$ at the $pp$ center-of-mass energy $E$ is defined as
\begin{equation}
S(E) = \sigma(E) \,E \, e^{2\pi\eta}\,, 
\end{equation}
where $\eta = \sqrt{m_p/E}\,\alpha/2$ is the Sommerfeld parameter, $m_p$ is the proton mass, $\alpha=1/137.036$ is the fine-structure constant, and $\sigma(E)$ is the $pp$ fusion cross section at energy $E$, which can be written as 
\begin{align}
 \label{eq:crossec}
 \sigma(E) = \int
  & \frac{\mathrm{d}^3p_e}{(2\pi)^3}\frac{\mathrm{d}^3p_\nu}{(2\pi)^3}\frac{1}{2E_e} \frac{1}{2E_\nu}
  \, 2\pi\delta\left(E+2m_p-m_d-\frac{q^2}{2m_d}-E_e-E_\nu\right)\nonumber\\
  & \frac{1}{v_{rel}}\,F(Z,E_e)\,\frac{1}{4}\sum \vert \langle f \vert \hat H_W \vert i \rangle  \vert^2\,.
\end{align}
Here $p_{e,\nu}$ are the positron and neutrino momenta, $E_{e,\nu}$
their energies, $m_d$ is the deuteron mass, $v_{rel}$ is the $pp$
relative velocity, and $q$ is the momentum of the recoiling
deuteron. The function $F(Z,E_e)$ accounts for the distortion of the
positron wave function due to the Coulomb field of the deuteron. Its
classical expression, which can be found in
Ref.~\cite{Feenberg1950399}, needs to be augmented with radiative
corrections. We enhance $F(Z,E_e)$ by 1.62\% to account for the
$\gamma W$ box diagram involving one nucleon, which has been
explicitly evaluated in Ref.~\cite{Kurylov:2002vj}, and ignore the
diagram involving both nucleons, which has not yet been
calculated. The summation in Eq.~\eqref{eq:crossec} runs over the spin
projections of all the initial- and the final-state particles. The
initial state $\vert i \rangle$ and the final state $\vert f \rangle$
are products of leptonic and nuclear states and the weak interaction
Hamiltonian $\hat H_W$ can be written in terms of the leptonic weak
current $j^\mu$ and the nuclear weak current $J^\mu$ as
\begin{equation}
\label{eq:weak_current_operator}
 \hat H_W = \frac{G_V}{\sqrt{2}}\int\mathrm{d}^3x \left[j^\mu(\mathbf{x}) {J_\mu}^\dagger(\mathbf{x})+\mbox{h.c.}\right],
\end{equation}
where $G_V$ is the vector coupling constant. Unless otherwise stated, the calculations presented below 
 adopt the value $1.149589\times10^{-11}\,\mathrm{MeV}^{-2}$, that corresponds to the
CKM matrix element $V_{ud}=0.9737$ and the process independent
radiative corrections $\Delta_R^V=0.02454$ as obtained in a recent
reanalysis of the superallowed beta decay rates~\cite{Hardy:2020} 
 (see Ref.~\cite{Feng:2020zdc} for a direct computation in Lattice QCD 
and Ref.~\cite{Seng:2020wjq} for a prediction with minimal phenomenological input).

The matrix element of the leptonic weak current operator $j^\mu$
between the leptonic wave functions can be obtained from Dirac algebra
and we refer the reader to Refs.~\cite{Marcucci:2013tda,Acharya:2016kfl} for details of
the derivation of the matrix elements of the nuclear weak current
operator $J^\mu$ between the nuclear wave functions. 
 In particular, we regularize the currents using the same 
Gaussian regulators that were  used in Refs.~\cite{Marcucci:2013tda,Acharya:2016kfl}. 
The incoming $pp$
system can be considered exclusively in an $s$-wave, as it has been
shown that higher partial-wave channels are suppressed by several
orders of magnitude (see Ref.~\cite{Acharya:2019zil} and also the
erratum of Ref.~\cite{Marcucci:2013tda}). Then, the leading
contribution to $J^\mu$ comes from the one-body (1B) Gamow-Teller (GT)
operator and the leading two-body (2B) corrections occur at
$\mathcal{O}\left([Q/\Lambda_b]^3\right)$ relative to 1B, which
correspond to next-to-next-to-leading order (NNLO) in the chiral
expansion.  It should be noted that these differ slightly from the
currents used in Ref.~\cite{Marcucci:2013tda}: besides the
leading-order GT 1B contribution, relativistic $1/m_N^2$ corrections to the 1B
GT term, where $m_N$ is the nucleon mass, were also included since they are
$\mathcal{O}\left([Q/\Lambda_b]^2\right)$ in the power counting
adopted in that work. In order to facilitate comparison with Ref.~\cite{Marcucci:2013tda}, we retain these terms in our updates to the study of Ref.~\cite{Marcucci:2013tda}  (model D discussed below) although they are small. The 2B current we use are the same as the ones used in Refs.~\cite{Marcucci:2013tda,Acharya:2016kfl}, but with the corrected~\cite{PhysRevLett.121.049901} relation (see also Refs.~\cite{PhysRevLett.103.102502,PhysRevLett.122.029901}) to the three-nucleon force parameter $c_D$ as discussed in Sec.~\ref{sec:cddr}. Unless otherwise stated, we use
the latest Particle Data Group (PDG) recommended value of the axial
coupling constant $g_A=1.2754\pm0.0013$~\cite{Workman:2022ynf} in the current operator in the 
calculations presented below. 
 
The nuclear wave functions are calculated either by solving the
Lippmann-Schwinger (LS) equations as in Ref.~\cite{Acharya:2016kfl},
or by applying the variational method (VM) of
Refs.~\cite{PhysRevLett.108.052502,Marcucci:2013tda}.  The chiral EFT
Hamiltonians include nuclear as well as electromagnetic potentials. In
this work, relativistic and radiative corrections to the Coulomb
interaction, discussed in Ref.~\cite{Carlsson:2015vda}, are included
explicitly in the calculation of the wave function, as in
Ref.~\cite{Marcucci:2013tda}.  This is in contrast to
Ref.~\cite{Acharya:2016kfl}, where we first calculated $S(E)$ using
$pp$ wave functions with nuclear plus Coulomb interaction and applied
a phenomenological correction to take such higher-order
electromagnetic effects into account. Finally, in
Ref.~\cite{Acharya:2016kfl}, we used in Eq.~\eqref{eq:crossec} the
deuteron mass $m_d$ obtained from the calculated value of the binding
energy, which gave a negligible numerical error since all of the
interactions used in Ref.~\cite{Acharya:2016kfl} were fit to the
experimental $^2$H binding energy. As we discuss below, since some of
the interactions we use in this work do not reproduce this energy very
well, we use the experimental value of
$m_d = 1875.61294257$~MeV~\cite{CODATA2018} instead.  Note that for
the interaction models which properly reproduce the $^2$H binding
energy, the theoretical and experimental value of $m_d$ obviously
coincide to sufficient precision.

\subsection{Relationship between the axial current and the the three-nucleon force}
\label{sec:cddr}

The axial 2B current contains a counterterm $\hat d_R$ that needs to be fixed 
before predictive calculations of $S(E)$ can be performed. This LEC has not yet been determined
from $A<3$ observables. In chiral EFT, fitting to $A>2$ data involves the relationship between $\hat d_R$, the $\pi N$ LECs $c_{3,4}$ and the pion-exchange part of the $3N$ force parameter $c_D$:
\begin{equation}
\label{eq:cddr}
\hat d_R = -\textstyle{\frac{m_N }{4g_A\Lambda_\chi}}c_D
+\frac{m_N}{3}c_3 + \frac{2m_N}{3}c_4 +\frac{1}{6}\,,  
\end{equation}
where $\Lambda_\chi \approx 700$~MeV is the breakdown
scale of chiral perturbation theory. With $c_{3,4}$ values constrained by $\pi N$ and/or $NN$ data, 
the LEC $\hat d_R$ (or $c_D$) can then be obtained by 
fitting it, along with the contact $3N$ force parameter $c_E$, to 
observables such as $^3$H $\beta$ decay rate and binding energy.

Following the suggestion of Ref.~\cite{gardestig:2006},  
Eq.~\eqref{eq:cddr} was first derived in Ref.~\cite{PhysRevLett.103.102502}, 
albeit with an incorrect factor in front of the $c_D$ term (the factor -1/4 was missing). 
This error, which propagated widely in the literature and also entered the results of 
Refs.~\cite{Marcucci:2013tda,Acharya:2016kfl}, was first corrected by Ref.~\cite{PhysRevLett.121.049901}. The corrected relation was used by Ref.~\cite{Acharya:2018qzk} to compute $S(0)$. The value $S(0) = 4.058 \times 10^{-23}$MeV~fm$^2$ quoted by Ref.~\cite{Acharya:2018qzk} was obtained using $^3$H $\beta$ decay rate to constrain $\hat d_R$, with a set of chiral interactions not explored in this work. Note that the main goal of Ref.~\cite{Acharya:2018qzk} was to compute the muon-deuteron capture rate $\Gamma_{\mu d}$ and not $S(E)$---no sources of theory errors on $S(0)$ were thoroughly examined other than the uncertainty in the $\Gamma_{\mu d} - S(0)$ correlation due to nucleon axial form factor.

Eq.~\eqref{eq:cddr} also enables us to constrain the axial current from strong-interaction observables. Therefore, different sets of experimental observables have been used, e.g., Ref.~\cite{lynn:2016} used ${}^4$He binding energy and elastic ${}^4$He-$n$ scattering data whereas Ref.~\cite{Maris:2020qne} used ${}^3$H binding energy and elastic ${}^3$H-$n$ scattering data.

\section{Results}
\label{sec:results}

We now present our results for $S(E)$ obtained using various chiral EFT interactions:  (i) model A which uses wave functions obtained by solving the Lippmann-Schwinger equation using a modern potential that is well adapted to 
EFT truncation studies because it is formulated at different orders, (ii) model B which uses the interactions of Ref.~\cite{Carlsson:2015vda} with the updated fits~\cite{Acharya:2018qzk} to account for the corrected $c_D$-$\hat d_R$ relation, (iii) model C where the statistical uncertainty of fitting $c_D$ to few-body observables has been studied in detail, and (iv) model D which uses the variational method of Refs.~\cite{Marcucci:2013tda,PhysRevLett.108.052502} along with interactions of Ref.~\cite{Entem:2003ft}, which requires a new fit of $c_D$. In Table~\ref{tab:models}, we summarize the salient features of these four models. 
\begin{table}[t]
  \begin{tabular}{c|c|c|c|c|c}
    Model & Method & $1/m_N^2$ currents & $g_A$ & $c_D$ & B(${}^2$H)     \\
    \hline
A & LS & excluded & 1.2754 & -1.626& 2.22038 \\
    \hline
B & LS & excluded & 1.2754 & see text &2.224$^{(+0)}_{(-1)}$~\cite{Carlsson:2015vda} \\
    \hline
C & LS & excluded & 1.2724 & -0.0047& 2.18553   \\
    \hline
D & VM & included & 1.2754 & see Table~\ref{tab:GTexp-newfit} & 2.22458 \\
    \hline
    \\
  \end{tabular}
  \caption{Main
      features of the four models adopted in this work. In particular,
      for each model we indicate whether the $1/m_N^2$ one-body currents
      are included, and we provide the adopted values for the single-nucleon
      axial coupling constant $g_A$, the LEC $c_D$, and  the calculated deuteron
      binding energy B($^2$H) in MeV.
    }
  \label{tab:models}
\end{table}

\subsection{Model A: the SMS-RS predictions}

Table~\ref{tab:order-by-order} shows the threshold values and energy-derivatives of 
$S(E)$ calculated using the SMS-RS potential of
Ref.~\cite{Reinert:2017usi} at regulator cutoff $\Lambda=500$~MeV and
the LS equations as in Ref.~\cite{Acharya:2016kfl}. In contrast to
older momentum-space chiral
interactions~\cite{Epelbaum:2003gr,Epelbaum:2003xx,Entem:2003ft,PhysRevC.91.051301,PhysRevLett.110.192502,Entem:2017gor}
that employed the conventional non-local Gaussian regulators along
with an additional spectral function
regularization~\cite{Kaiser:1997mw} in the two-pion exchange diagrams,
the potential of Ref.~\cite{Reinert:2017usi} uses a ``semi-local''
regularization scheme which consists of a local regulator for the
pion-exchange parts and non-local Gaussian regulator for the contact
parts of the $NN$ interaction in order to preserve the long-range
parts that are unambiguously determined in chiral EFT. The subleading
$\pi N$ LECs in this potential have been fixed to the precise values
obtained from a Roy-Steiner analysis \cite{Hoferichter:2015tha} of the
$\pi N$ scattering data.  The nucleon-nucleon ($NN$) LECs in the
SMS-RS potential have been fit to mutually consistent $np$ and $pp$
scattering data of the Granada 2013
database~\cite{NavarroPerez:2013mvd}.  In the 2B current, we use
$\hat d_R$ given by Eq.~\eqref{eq:cddr} with the value of $c_D=-1.626$
obtained in Ref.~\cite{Maris:2020qne} by fitting to $Nd$ scattering
cross section data; however, we note that the theoretical
uncertainties in the estimation of $c_D$ were not explored in that
reference. It should be remarked that, thanks to the availability of the SMS-RS
potential at various orders in the EFT expansion,
it is possible to perform an order-by-order calculation consistently for
both potential and current. This is different from the study of
Ref.~\cite{Marcucci:2013tda}, where the nuclear interaction order was
fixed at N3LO, while the different orders were considered only for the
chiral expansion of the axial current.

We evaluate $S(E)$ in the energy interval $E = [1,30]$~keV and fit a 
polynomial in $E$ to obtain the threshold $S$-factor
$S(0)$ and its derivatives. The extracted values depend on the 
degree of the fitted polynomial. We obtained stable values, particularly for 
$S(0)$ and $S^\prime(0)$, for third, fourth and fifth degree polynomials. 
Unless otherwise stated, we quote values from cubic fits. 

\begin{table}[t]
        \begin{tabular*}{0.95\textwidth}{@{\extracolsep{\fill}} lcccc}
          Order & $S(0)$ [MeV fm$^2$] & $S'(0)/S(0)$ [MeV$^{-1}$]
          & $S''(0)/S(0)$ [MeV$^{-2}$] & $S'''(0)/S(0)$ [MeV$^{-3}$]\\
          \hline
                LO                 & 4.143 $\times10^{-23}$  & 10.75 & 306.75 & -5150 \\
                NLO                 & 4.094 $\times10^{-23}$ & 10.81 & 312.78 & -5370\\
                NNLO           & 4.100 $\times10^{-23}$ & 10.83 & 313.72 & -5382 \\
          \hline
        \end{tabular*}
        \caption{The threshold $S$-factor $S(0)$, and the first and
          second derivatives of the $S$-factor at threshold divided by
          $S(0)$ in units of MeV fm$^2$, MeV$^{-1}$, MeV$^{-2}$ and
          MeV$^{-3}$, respectively, at different orders of the chiral EFT
          expansion for both the nuclear interaction and
            current. The SMS-RS interaction is used. }
        \label{tab:order-by-order}
\end{table}

We assess the uncertainty of the
$S$-factor due to the EFT truncation by following the method described
in Refs.~\cite{PhysRevC.92.024005,PhysRevLett.115.122301}. We first
express the $S$-factor as
\begin{equation}
S(0) = S_{\rm LO}\sum_{n=0}^3 c_n\left( \frac{Q}{\Lambda_b}\right)^n~,
\end{equation}
where $S_{\rm LO}$ denotes the leading order (LO) result for the
$S$-factor given in Table~\ref{tab:order-by-order}, $Q$ denotes the
inherent momentum scale of the problem, and $\Lambda_b$ is the
breakdown scale. An estimate of the truncation error is then obtained
by calculating $(Q/\Lambda_b )^4 \max(|c_0 |, |c_2 |, |c_3 |)$. Using
the order-by-order results of Table~\ref{tab:order-by-order} to obtain
the expansion coefficients $c_n$, the pion mass for the momentum scale
$Q$ and a conservative estimate of $\Lambda_b = 500$~MeV, we obtain an
uncertainty 0.024$\times 10^{-23}$~MeV~fm$^2$ for the SMS-RS interaction, leading to
the prediction
$S(0)=(4.100\pm0.024\mathrm{(syst)})\times10^{-23}~\rm{MeV\,fm}^2$. Here
we have used the label ``syst'' to emphasize that is is an estimate of
the systematic uncertainty from the truncation of the EFT
expansion. We note that 
this estimate has been found to roughly
correspond to a 68\% Bayesian credible interval for a particular
choice of Bayesian priors for the expansion coefficients
$c_n$~\cite{PhysRevC.92.024005}.

\subsection{Model B: the NNLOsim updates}

In Ref.~\cite{Acharya:2016kfl}, we employed the NNLOsim family of 42
interactions~\cite{Carlsson:2015vda} to assess the uncertainty in the
$pp$ fusion rate due to the statistical uncertainties in the LECs, the
systematic uncertainty due to the chiral EFT cutoff dependence, and
the systematic variations in the database used to calibrate the $NN$
interaction.  At each of the 7 different cutoff values
$\Lambda=[450, 475, 500, 525, 550, 575, 600]$~MeV, the 26 LECs in the
$NN$ and $\pi N$ sectors were {simultaneously} fit to 6 different
pools of input data, leading to 42 interactions.  Specifically, the
input data consisted of $NN$ scattering data at different truncations
of the maximum scattering energy $T_\mathrm{lab}^\mathrm{max}$,
$\pi N$ cross sections, the binding energies and charge radii of
$^{2,3}$H and $^3$He, the one-body quadrupole moment of $^2$H, as well
as the comparative $\beta$-decay half-life of $^3$H. The NNLOsim
interactions have been refit~\cite{Acharya:2018qzk} to account for the
update in the $c_D$-$\hat d_R$ relation given in
Eq.~\eqref{eq:cddr}~\cite{PhysRevLett.121.049901}, which enters the
calculation of the $^3$H GT matrix element fit to tritium
$\beta$-decay. The resulting values of $c_D$ for this model are roughly
uniformly distributed between -2 and 2. We obtain
$4.091\times10^{-23}~\rm{MeV\,fm}^2$ for the average $S(0)$ value for
the refitted NNLOsim interactions. This is an upward revision, by
about $1.1\%$, partly driven by the refitting of $c_D$
  but mainly by the change in the $g_A$ value recommended by the
  PDG~\cite{Workman:2022ynf}, from the result
$S(0)=4.047^{+0.024}_{-0.032}\times10^{-23}~\rm{MeV\,fm}^2$ obtained
in Ref.~\cite{Acharya:2016kfl}, where the quoted uncertainty mainly
reflected the sensitivity of the $S$-factor to input data used to
calibrate the LECs of the EFT and to the short-distance behavior of
the $NN$ interactions.  While a full reanalysis of the
  uncertainty estimates of Ref.~\cite{Acharya:2016kfl} is beyond the
  scope of this work, we note that $S(0)$ spans the range
  4.081$\times 10^{-23}$ to 4.095~$\times 10^{-23}$~MeV~fm$^2$ for the
  42 interactions. Unlike Ref.~\cite{Acharya:2016kfl}, this range does
  not include the statistical fitting errors of the LECs. It is also
  instructive to compare the EFT truncation uncertainty obtained above
  using SMS-RS interaction to a similar estimate using one of the 42
  interactions from the NNLOsim family. To this end, we choose the
  interaction with $T_\mathrm{lab}^\mathrm{max} = 290$~MeV and
  $\Lambda=500$~MeV, and obtain $13.537\times10^{-23}$,
  $4.869\times10^{-23}$ and $4.092\times10^{-23}~\rm{MeV\,fm}^2$ for
  $S(0)$ at LO, NLO and NNLO. Using the procedure discussed above to
assess the EFT truncation error, we obtain
$S(0)=(4.092\pm0.178\mathrm{(syst)})\times10^{-23}~\rm{MeV\,fm}^2$. We
note that this large uncertainty is a result of the LO value being
rather large because the deuteron properties are not well reproduced
by this interaction at this order. We therefore consider the estimate
of the systematic EFT truncation error obtained from the SMS-RS
interaction to be more reliable for this problem.

\subsection{Model C: uncertainty in calibrating the axial contact current from few-body observables}

An additional uncertainty that has not been included in the SMS-RS
result quoted above,
$S(0)=(4.100\pm0.019\mathrm{(syst)})\times10^{-23}~\rm{MeV\,fm}^2$, is
the uncertainty arising from the axial 2B contact current with
$\hat d_R$ determined using Eq.~\eqref{eq:cddr}. To estimate this, we
consider the Bayesian posterior probability distribution function
(PDF) of $c_D$-$c_E$ obtained by Ref.~\cite{Wesolowski:2021cni}. In
that study, as in the SMS-RS interaction, the $\pi N$ LECs were fixed
to central values determined by the Roy-Steiner analysis in
Ref.~\cite{Hoferichter:2015tha}. The $NN$ LECs at LO, NLO, and NNLO
were then fixed by performing a new fit to the $np$ and $pp$
scattering data with $T_\mathrm{lab}^\mathrm{max}=290$~MeV gathered
from the Granada 2013 database~\cite{NavarroPerez:2013mvd} as well as
the empirical $nn$ effective range parameters. Constraints on $c_D$
and $c_E$ were obtained from fitting to binding energies of $^3$H and
$^4$He, the charge radius of $^4$He, and the GT matrix element of the
$^3$H extracted from tritium $\beta$-decay followed by marginalization
of all other parameters with both the uncertainty of the chiral EFT
Hamiltonian and the uncertainties in the experimental measurements
taken into account. The experimental value of the $^3$H $\beta$ decay half life $(1129.6\pm 3)$~s~\cite{Akulov:2005} 
adopted by Ref.~\cite{Wesolowski:2021cni} corresponds to our Fit-2 discussed below.

It was found that, at the older PDG value of
$g_A=1.2724$~\cite{PhysRevD.98.030001}, the joint PDF of $c_D$-$c_E$
was well described by a multivariate $t$-distribution $t_\nu(m,S)$
with $\nu \approx 2.8$ degrees of freedom, a mean vector
$m = [-0.0047, -0.1892]$ and a scale matrix of
\begin{equation}
    S =  \begin{bmatrix}
0.250 & 0.043 \\
0.043 & 0.008 
\end{bmatrix}
\end{equation}
at one standard deviation ($1\sigma$). At least for the energy range
considered in this work, $S(E)$ has approximately linear dependence on
the values of $c_D$ within the $1\sigma$ range for this joint PDF. The
uncertainty on $S(E)$ can therefore be very well approximated by
sampling $c_D$ from its marginal $t$-distribution, which spans the
approximate $1\sigma$ range $c_D \in [-0.615,0.615]$, and computing
$S(E)$ at the $1\sigma$ margins. This gives
$S(0) = (4.155 \pm 0.013(\mathrm{stat}))\times10^{-23}$~MeV~fm$^2$. Here we used the
label ``stat" to indicate that this is an estimate of the statistical
uncertainty from the $c_D$ probability distribution. The rather large
$S(0)$, in spite of the smaller value of 1.2724 adopted for $g_A$, is
a result of slow convergence of the deuteron properties in this EFT
expansion up through NNLO. This indicates that the fitting procedure
adopted for this interaction results in large EFT truncation error
even at NNLO for extreme low-energy observables. Nevertheless, we
consider the quoted uncertainty from $c_D$ variation to be a
reasonable estimate of the uncertainty in $S(0)$ from fitting the
two-body contact axial current to few-body observables.

\subsection{Model D: updates to Idaho-N3LO results}

We now turn our attention to the study of the $pp$ fusion performed
using the VM to calculate the deuteron and $pp$ wave functions as in
Ref.~\cite{Marcucci:2013tda} from the Idaho-N3LO potential with
$\Lambda=500$ MeV~\cite{Entem:2003ft}. With respect to
Ref.~\cite{Marcucci:2013tda}, we have put particular attention to the
convergence on the basis expansion for the deuteron wave function, and
on the integration range used to calculate the transition operator
matrix element. In particular, we have verified that not only the
deuteron binding energy, but also the asymptotic normalization
constants and the so-called $D/S$-state ratio $\eta=A_D/A_S$ are well
reproduced. Furthermore, we have verified that the integral range of
$r_{max}=50$ fm is sufficient to reach convergence of the results. In
fact, by going from $r_{max}=50$ fm to $r_{max}=60$ fm, the change in
$S(0)$ is beyond the third decimal digit.  Therefore, the results we
are going to present are at convergence at least up to the the
third decimal digit of $S(0)$. We note that such an accuracy was not
the primary goal of the work of Ref.~\cite{Marcucci:2013tda}, where
the results were obtained with a theoretical accuracy of the order of
1\%.

In order to use the most recent values for the fundamental constants
entering the calculations, we, first of all, update the value for the
experimental GT matrix element in tritium $\beta$-decay
$\langle GT_{exp} \rangle$. Note that $\langle GT_{exp} \rangle$ and the $A=3$ binding energies were
the observables of choice used to fix the LEC $c_D$, and consequently
$\hat d_R$, and the LEC $c_E$, that parametrizes the three-nucleon
contact interaction entering at NNLO (see, for instance,
Ref.~\cite{PhysRevLett.108.052502}). The GT matrix element is
  defined as
\begin{equation}
  \langle GT\rangle^2=\bigg[\frac{2 ft_{0^+\rightarrow 0^+}}{ft_{^3{\rm H}}}
    -\langle F\rangle^2\bigg]\,\frac{1}{f\,g_A^2} \label{eq:GT2} \ ,
\end{equation}
where $g_A$ is the single-nucleon axial coupling constant,
$f=f_A/f_V=1.00529$ is the ratio of the axial and vector Fermi
functions, $ft_{^3{\rm H}}$ and $ft_{0^+\rightarrow 0^+}$ are the
$ft$-values for tritium $\beta$-decay and for superallowed
$0^+\rightarrow 0^+$ transitions. From $\langle GT\rangle$, we define
$\langle GT_{exp} \rangle$ as $\langle GT_{exp} \rangle={\langle GT\rangle}/{\sqrt{3}}$.  
With this definition, $\langle GT_{exp} \rangle$
  is related to the
  reduced matrix element of the electric dipole axial operator $E_1$
  used in Refs.~\cite{PhysRevLett.103.102502,Wesolowski:2021cni} via the relation
  $\langle GT_{exp} \rangle=\sqrt{\pi} E_1/g_A$. In the
present calculation, we have used $g_A=1.2754\pm 0.0013$, according to
the PDG, and $ft_{0^+\rightarrow 0^+}=(3072.24\pm 1.85)$ s, according
to Ref.~\cite{Hardy:2020}. This value is consistent with the value of
$G_V$ used in Eq.~(\ref{eq:weak_current_operator}). Furthermore, we
have used $\langle F \rangle^2=0.99967$, as it was obtained in
Refs.~\cite{PhysRevLett.108.052502} with the Idaho-N3LO
potentials. This value is quite different from the one obtained using
the phenomenological AV18/UIX interaction model, and used for instance
in Ref.~\cite{Schiavilla:1998je}, $\langle
F\rangle^2=0.9987$. However, we have verified that the impact on
$\langle GT_{exp} \rangle$ of this different $\langle F \rangle^2$ is
negligible. Finally, we have adopted for $ft_{^3{\rm H}}$ three
different values: $(1134.6\pm 3.1)$~s as obtained in
Ref.~\cite{Simpson:1987}, $(1129.6\pm 3)$~s  as obtained in
Ref.~\cite{Akulov:2005} and adopted by Model C above, and $(1132.1\pm 4.3)$~s as used already in
Ref.~\cite{PhysRevLett.108.052502} and obtained averaging these two 
values and summing the errors in quadrature. The $\langle GT_{exp} \rangle$
obtained with these three values for $ft_{^3{\rm H}}$ will be labelled
Fit-1, Fit-2 and Fit-3, respectively.  The three values for
$\langle GT_{exp} \rangle$, together with their uncertainties arising from the
experimental errors on $g_A$, $ft_{0^+\rightarrow 0^+}$ and
$ft_{^3{\rm H}}$, are listed in Table~\ref{tab:GTexp-newfit}. In the
table, we report also the corresponding ranges for $c_D$ obtained, as
mentioned above, with the fitting procedure of
Ref.~\cite{PhysRevLett.108.052502}.
\begin{table}[t]
  \begin{tabular}{c|c|c}
    & $\langle GT_{exp}\rangle$ & $c_D$ \\
    \hline
    Fit-1 & $0.9488\pm 0.0019$ & 0.1836--0.7680 \\
    Fit-2 & $0.9514\pm 0.0019$ & 0.3830--0.9690 \\
    Fit-3 & $0.9501\pm 0.0024$ & 0.2833--0.8685 \\
  \end{tabular}
  \caption{Values for $\langle GT_{exp}\rangle$ as obtained using
    three different values for $ft_{^3{\rm H}}$, $1134.6 \pm 3.1$ s
    labelled Fit-1, $1129.6\pm 3$ s labelled Fit-2, and $1132.1\pm 4.3$
    labelled Fit-3, respectively. The corresponding ranges for the LEC $c_D$
    obtained using the Idaho-N3LO potential with $\Lambda=500$ MeV 
    are also listed.}
  \label{tab:GTexp-newfit}
\end{table}
With the new values for the LEC $c_D$, or equivalently $\hat d_R$, we
present in Table~\ref{tab:VM-LSpp} the results for the zero-energy
astrophysical $S$-factor $S(0)$, and its first, second and third derivatives,
$S^\prime(0)/S(0)$,  $S^{\prime\prime}(0)/S(0)$ and $S^{\prime\prime\prime}(0)/S(0)$. We list in the
table also the results obtained with the same constants and the same
potential using the LS equation and the axial current of
Ref.~\cite{Acharya:2016kfl}. By inspecting the values in the table, we can
conclude that (i) the VM and LS calculations are in very good
agreement as the small differences are consistent with the anticipated size of the 
relativistic corrections to the GT 1B operator which were neglected in the LS calculations
 but were retained in the VM calculations;
(ii) the dependence on the adopted value for $ft_{^3{\rm H}}$ is very
weak; (iii) the theoretical uncertainty arising from the range of
$c_D$ values allowed for each fitting procedure, and therefore
ultimately related to the experimental error on $\langle GT_{exp} \rangle$, is
extremely small for $S(0)$ and $S^{\prime\prime}(0)/S(0)$, and beyond
the second decimal digit for $S^\prime(0)/S(0)$; (iv) the calculated
$S(0)$ is in excellent agreement with the $S(0)$ value obtained with
the SMS-RS and NNLOsim interactions.
(v) The value for $S(0)$ is $\sim 2.2$\% larger than the value of
Ref.~\cite{Marcucci:2013tda}. The difference arises from the improved
convergence of the nuclear wave functions adopted in this work, as
well as from the increased values for the single-nucleon axial
coupling constant $g_A$. 

\begin{table}[t]
  \begin{tabular}{c|c|c|cccc}
    & Method & $1/m_N^2$ currents & $~~~~S(0)~~~~$ & $~~S^\prime(0)/S(0)~~$ & $~~S^{\prime\prime}(0)/S(0)~~$ & $~~S'''(0)/S(0)~~$\\
    \hline
    Fit-1 & VM & included & 4.115(4) & 10.60 & 347.1 & -6908 \\
          & LS & excluded & 4.101(4) & 10.83 & 313.8 & -5382\\
    \hline
    Fit-2 & VM &  included & 4.118(4) & 10.60 & 347.1 & -6907 \\
          & LS &  excluded & 4.104(4)& 10.83 & 313.8 & -5381\\
    \hline
    Fit-3 & VM &included & 4.117(4) & 10.60 & 347.1 & -6908\\
          & LS & excluded & 4.104(4) & 10.83 & 313.8 & -5382\\
    \hline
  \end{tabular}
  \caption{Values for the zero-energy astrophysical $S$-factor $S(0)$
    (in $10^{-23}$ MeV fm$^2$), its first, second and third derivatives
    $S^\prime(0)/S(0)$ (in MeV$^{-1}$), 
    $S^{\prime\prime}(0)/S(0)$ (in MeV$^{-2}$) 
    and $S^{\prime\prime\prime}(0)/S(0)$
    (in MeV$^{-3}$) obtained with the Idaho-N3LO potential with
    $\Lambda=500$ MeV, using the variational method (VM) of
    Ref.~\cite{Marcucci:2013tda}, or the Lippmann-Schwinger (LS)
    equation as in Ref.~\cite{Acharya:2016kfl}. The results labelled
    Fit-1, Fit-2, and Fit-3 are obtained consistently with the values
    of $c_D$ listed in Table~\ref{tab:GTexp-newfit}. The numbers in
    parentheses are the theoretical error arising from the range of
    $c_D$ values allowed for each fitting procedure. For
    $S^\prime(0)/S(0)$ these errors are beyond the quoted digits.}
  \label{tab:VM-LSpp}
\end{table}

\section{Summary and Outlook}
\label{sec:summary}
We calculated the proton-proton fusion rate using
various chiral interactions: the Idaho-N3LO interaction 
at regulator cutoff of 500~MeV~\cite{Entem:2003ft}, the 
NNLOsim family of 42 interactions fit
at 7 different regulator cutoffs to 6 different pools of input data, 
and two additional sets of interactions---one
non-local~\cite{Wesolowski:2021cni} and another
``semi-local''~\cite{Reinert:2017usi}---in which the pion-nucleon
coupling constants are fixed to precise values determined from
Roy-Steiner analysis. We obtained an estimate for the uncertainty from
the truncation of the EFT expansion and also assessed the uncertainty
from fixing the LEC $\hat d_R$ in the two-body axial current using
$c_D$ fitted to $A\geq3$ data. Furthermore, by using the Idaho-N3LO
potential, we have performed a benchmark calculation between the two
different approaches first applied to the $pp$ reaction in
Refs.~\cite{Marcucci:2013tda} and~\cite{Acharya:2016kfl}. 
We have used recent values of the fundamental constants and have 
used the corrected relationship between the axial current LEC $\hat d_R$ 
and the three-body force parameter $c_D$. 

The threshold $S$-factor $S(0)$ obtained from the various EFT
interactions indicate an upward revision from the recommendation made
by Ref.~\cite{Adelberger:2010qa}. It is remarkable that the $S$-factor
value obtained using the SMS-RS interactions~\cite{Reinert:2017usi} is
in excellent agreement with the NNLOsim result although its $c_D$
value was obtained without fitting to electroweak data. We also show
that variation in $c_D$ within the 68\% Bayesian credible interval
obtained from the joint $c_D-c_E$ distribution calculated in
Ref.~\cite{Wesolowski:2021cni} translates to an uncertainty of
$0.013~\mathrm{MeV}~\mathrm{fm}^2$ in $S(0)$, which is smaller than
the uncertainty of $0.019~\mathrm{MeV}~\mathrm{fm}^2$ stemming from
the truncation of the chiral expansion in the SMS-RS potential. We do
note, however, that the potential of Ref.~\cite{Wesolowski:2021cni}
underbinds the deuteron at all orders (by about 75\% at LO, 15\% at
NLO and 1\% at NNLO) which causes $S(E)$ to be overpredicted, even
with just one-body current. We encounter this issue also at LO and NLO
in the NNLOsim interactions, which again impacts the extraction of
truncation error. This highlights the importance of ensuring that the
low-energy properties of the $NN$ system, {e.g.} the binding energy
and the asymptotic normalization coefficient of the deuteron, are
reasonably well reproduced while fitting the LECs of chiral EFT. Given
the difficulty in assessing the truncation error of the EFT expansion
due to slow order-by-order convergence in the non-local chiral
interactions used in this work, we consider the central value obtained
using the SMS-RS interaction to be more reliable and the corresponding
truncation error estimate to be better calibrated. Combining the
systematic EFT truncation error with the statistical uncertainty that
arises from the PDF of $c_D$ and the range obtained by varying $g_A$
within the PDG recommendation of $1.2754\pm0.0013$, we recommend
\begin{equation}
\label{eq:recommendation}
S(0) = (4.100\pm0.024\mathrm{(syst)}\pm0.013\mathrm{(stat)}\pm0.008(g_A))\times10^{-23}~\mathrm{MeV\,fm}^2\,. 
\end{equation}
Note that this value is consistent with the ones obtained using the
NNLOsim and Idaho-N3LO interactions, and with the value
  $S(0)=(4.14\pm0.01\pm0.005\pm0.06)\times10^{-23}~\mathrm{MeV~fm}^2$
  recently obtained by De-Leon and Gazit in Pionless
  EFT~\cite{De-Leon:2022omx}.  Furthermore, we find with the SMS-RS
potential $S^\prime(0)/S(0)$ and $S^{\prime\prime}(0)/S(0)$ values of
$10.83~\mathrm{MeV}^{-1}$ and $313.72~\mathrm{MeV}^{-2}$,
respectively.  The values of these energy derivatives of $S(E)$ depend
on the interaction used, calculation method and degree of the fitted
polynomial.  We refrain from performing a detailed analysis of
uncertainties for them as they are not relevant at currently
achievable precision in the energy range of interest for solar
conditions.

Finally, we would like to remark that the $S$-factor is not accurately
predicted by the chiral EFT interactions of
Refs.~\cite{Carlsson:2015vda,Wesolowski:2021cni} at low chiral orders
if the deuteron bound-state properties are not adequately
reproduced. This highlights the importance of calibrating the
interactions by fitting to experimental data using strategies that
also lead to systematic convergence pattern for very low-energy
observables such as $pp$ fusion at solar conditions. Bayesian model
mixing between chiral and pionless EFT to ensure that the deuteron and
the $pp$ effective-range parameters are well reproduced appears to be
a promising strategy for reliably predicting solar $pp$ fusion
rate. Furthermore, we believe that a thorough study of the $pp$ fusion reaction 
should be performed using the largest possible variety of chiral EFT
interactions, local or non-local, regularized in co-ordinate or
momentum space, with or without the inclusion of $\Delta$-isobar
degrees of freedom, possibly available a different chiral orders, as,
for instance, those of
Refs.~\cite{Piarulli:2014bda,Nosyk:2021pxb,Entem:2017gor,Saha:2022oep}. Such
a study is beyond the scope of this work, but is definitely highly
recommended. Work along this line, on the footsteps of
Refs.~\cite{Ceccarelli:2022cpz,Gnech:2023pc}, is currently underway.

\section*{Acknowledgments}
We would like to thank Andreas Ekstr\"om, Rocco Schiavilla and Michele
Viviani for useful suggestions and comments. This work has been
supported by the National Science Foundation under Grant
Nos. PHY-1555030 and PHY-2111426, by the Office of Nuclear Physics,
U.S. Department of Energy under Contract No. DE-AC05-00OR22725, and by
the Office of High Energy Physics, U.S. Department of Energy under
Contract No. DE-AC02-07CH11359 through the Neutrino Theory Network
Fellowship awarded to BA.

\end{document}